\documentclass[runningheads]{llncs}
\usepackage[hyphens]{url}
\usepackage{hyperref}
\hypersetup{breaklinks=true}
\usepackage{graphicx}
\usepackage[backend=biber, style=alphabetic, sorting=ynt]{biblatex}
\usepackage[inline]{enumitem}
\usepackage{microtype}
\usepackage{xurl}
\usepackage{pdflscape}
\usepackage[T1]{fontenc}
\usepackage{lmodern}
\usepackage{csquotes}

\usepackage{adjustbox}
\usepackage{graphicx}

\usepackage{amsmath}
\usepackage{amssymb}

\usepackage{booktabs}
\usepackage{tabularx}
\usepackage{multirow}
\usepackage{makecell}
\usepackage[table]{xcolor} 

\usepackage{verbatim}

\title{Reuse of Public Keys Across UTXO and Account-Based Cryptocurrencies}
\author{
Rainer St\"utz\inst{1} \and
Nicholas Stifter\inst{2} \and
Melitta Dragaschnig\inst{3} \and
Bernhard Haslhofer\inst{1,5} \and
Aljosha Judmayer\inst{4}
}
\authorrunning{St\"utz et al.}

\institute{
Complexity Science Hub
\email{\{stuetz, haslhofer\}@csh.ac.at}
\and
SBA Research
\email{nstifter@sba-research.org}
\and
AIT Austrian Institute of Technology
\email{melitta.dragaschnig@ait.ac.at}
\and
University of Vienna
\email{aljosha.judmayer@univie.ac.at}
\and
Iknaio Cryptoasset Analytics GmbH
}

\date{\today}

\newcommand\numPtoPK{1,271,869}
\newcommand\numPtoMS{1,434,621}
\newcommand\numReusedKeys{1,604,614}
\newcommand\numActiveReusedKeys{1,429,008}

\newcommand\numUniqueEth{226,017,011}
\newcommand\numReusedKeysUtxoAccount{831,056}
\newcommand\numReusedKeysBtcETH{497,178}

\newcommand\numReusedKeysAVAX{1,872}
\newcommand\numReusedTxnsAVAX{5,138}

\newcommand\numUniqueTrx{266,333,226}

\newcommand\numTrxEthPassive{717,287}
\newcommand\numTrxEthActive{654,032}
\newcommand\numTrxEthDelta{63,255}
\newcommand\numTrxEthActivePercent{91}
\newcommand\numTrxEthPassivePercentofAddresses{0.2}
\newcommand\numTrxEthPassivePercentofPublicKeys{0.3}

\newcommand\numReuseHarrignAllLtcBtc{191,759}
\newcommand\numReusePkAllLtcBtc{170,680}
\newcommand\numPercentHarrignAllLtcBtc{89}

\newcommand\numReuseHarrignAllLtcDoge{28,276}
\newcommand\numReusePkAllLtcDoge{85,700}
\newcommand\numPercentHarrignAllLtcDoge{300}

\newcommand\numReuseHarrignAllBtcDoge{73,682}
\newcommand\numReusePkAllBtcDoge{153,378}
\newcommand\numPercentHarrignAllBtcDoge{200}

\newcommand\numReuseHarrignAllLtcBtcDoge{15,056}
\newcommand\numReusePkAllLtcBtcDoge{25,950}
\newcommand\numPercentHarrignAllLtcBtcDoge{170}

\newcommand\numReuseHarrignDogeLtc{2782}
\newcommand\numReusePkDogeLtc{2615}
\newcommand\numReusePkDogeLtcAll{2763}
\newcommand\numPercentHarrignDogeLtc{93}
\newcommand\numPercentHarrignDogeLtcAll{99}

\newcommand\numReuseHarrignBtcDoge{5371}
\newcommand\numReusePkBtcDoge{4767}
\newcommand\numReusePkBtcDogeAll{4923}
\newcommand\numPercentHarrignBtcDoge{89}
\newcommand\numPercentHarrignBtcDogeAll{92}

\newcommand\numReuseHarrignBtcLtc{36,209}
\newcommand\numReusePkBtcLtc{21,398}
\newcommand\numReusePkBtcLtcAll{25,182}
\newcommand\numPercentHarrignBtcLtc{59}
\newcommand\numPercentHarrignBtcLtcAll{70}

\newcommand\numReuseHarrignBtcLtcDoge{2139}
\newcommand\numReusePkBtcLtcDoge{2115}
\newcommand\numReusePkBtcLtcDogeAll{2203}
\newcommand\numPercentHarrignBtcLtcDoge{98}
\newcommand\numPercentHarrignBtcLtcDogeAll{103}

\newcommand\numBtcHeightHarrigan{520,650}
\newcommand\numLtcHeightHarrigan{1,413,308}
\newcommand\numDogeHeightHarrigan{2,200,704}
\newcommand\numBtcHeight{890,325}
\newcommand\numLtcHeight{2,871,702}
\newcommand\numDogeHeight{5,649,129}

\begin{document}

\maketitle
\setcounter{footnote}{0}

\begin{abstract}
It is well known that reusing cryptocurrency addresses undermines privacy.
This also applies if the same addresses are used in different
cryptocurrencies. Nevertheless, cross-chain address reuse appears to be a
recurring phenomenon, especially in EVM-based designs. Previous works
performed either direct address matching, or basic format conversion, to
identify such cases. However, seemingly incompatible address formats e.g., in
Bitcoin and Ethereum, can also be derived from the same public keys, since
they rely on the same cryptographic primitives. In this paper, we therefore
focus on the underlying public keys to discover reuse within, as well as
across, different cryptocurrency networks, enabling us to also match
incompatible address formats. Specifically, we analyze key reuse across
Bitcoin, Ethereum, Litecoin, Dogecoin, Zcash and Tron. Our results reveal that
cryptographic keys are extensively and actively reused across these networks,
negatively impacting both privacy and security of their users. We are hence
the first to expose and quantify cross-chain key reuse between UTXO and
account-based cryptocurrencies. Moreover, we devise novel clustering methods
across these different cryptocurrency networks that do not rely on heuristics
and instead link entities by their knowledge of the underlying secret key.
\end{abstract}

\section{Introduction}\label{sec:introduction}

It is well known that reusing cryptocurrency addresses, both within the same
cryptocurrency, as well as across different networks (cross-chain), undermines
the privacy of users and impacts
security~\cite{meiklejohn2013fistful,harrigan2018airdrops,lin2025connector}.
In this paper we explore a simple, yet powerful, technique to cluster entities
within and across different cryptocurrencies by focusing on the underlying
cryptographic keys used for signing transactions, rather than the addresses
which are derived from these keys. Hereby, we leverage the fact that many
cryptocurrencies rely on the same underlying cryptographic primitive, namely
ECDSA over \emph{secp256k1}~\cite{secg2010sec2}, for signing in a majority of
transactions.

Previous research has focused either on direct address matching or conversion
between compatible address formats to link key reuse within and across closely
related UTXO
cryptocurrencies~\cite{harrigan2018airdrops,kalodner2020blocksci,%
smolenkova2025keylinker} and, independently, within EVM-compatible account
models~\cite{zhang2022cltracer,lin2025connector}. However, by relying on
addresses rather than directly on the underlying public keys, these analyses
are limited to cryptocurrencies with compatible address/hash types. Crucially,
the address formats of the most prevalent designs for UTXO and account based
systems, i.e. Bitcoin and Ethereum, are not compatible although both use the
same digital signature scheme and curve for most transactions. Hence, until
now key reuse\footnote{For the most part address reuse also implies reuse of
the same cryptographic keys.} across these two different designs has not been
explored.

\begin{figure*}[ht]
  \centering
  \includegraphics[width=0.85\textwidth]{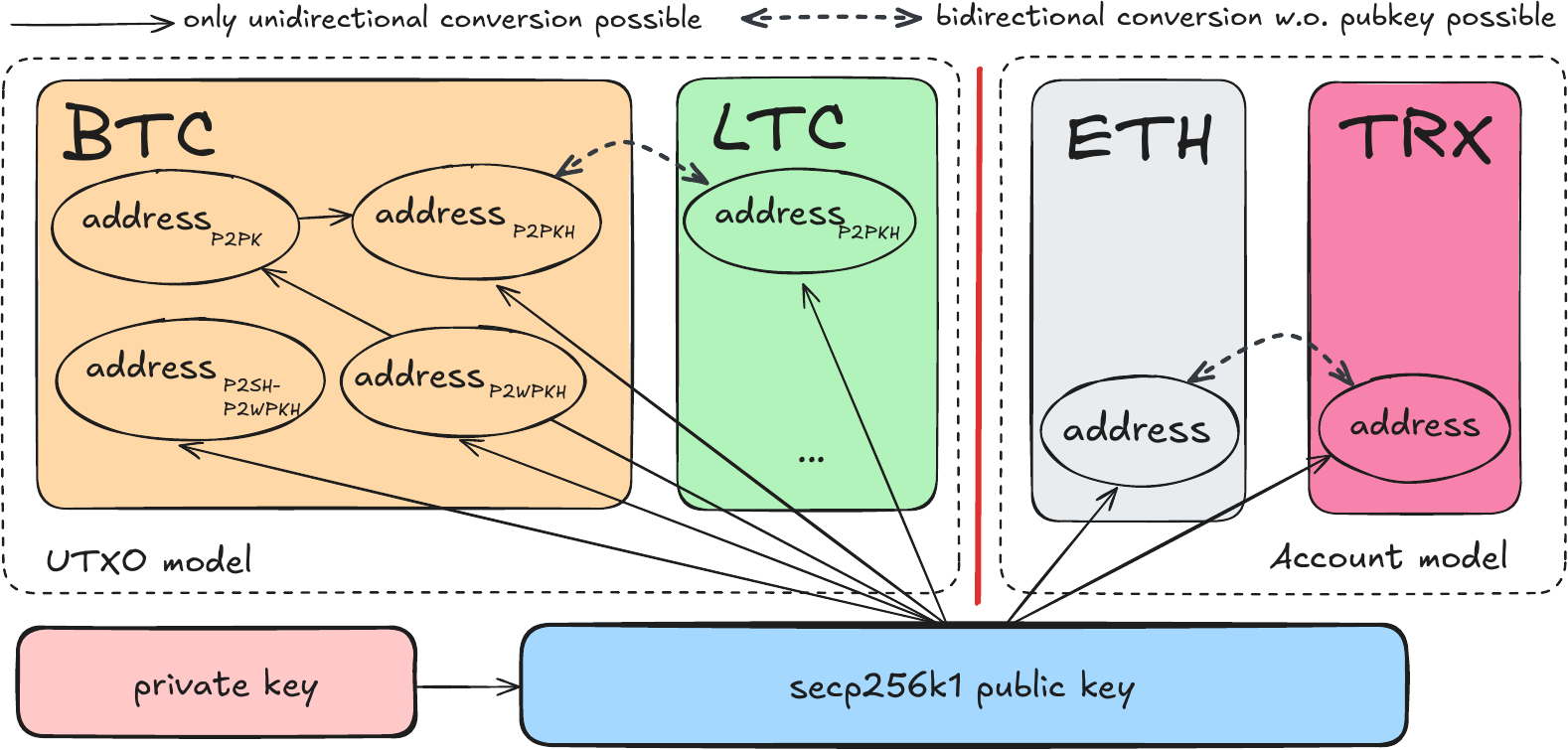}
  \caption{Address formats of different cryptocurrencies can be derived from
    the same public key. Some formats may be converted into each other without
    knowledge of the underlying public key while others can not.}
  \label{fig:overview}
\end{figure*}

\noindent
In this work we address this research gap by treating the public keys
associated with different cryptocurrency addresses as first-class citizens in
the context of key reuse. Knowledge of a public key allows us to generate a
one-to-many mapping (from public key to addresses) both within, as well as
across different cryptocurrency designs (see Figure~\ref{fig:overview} for
details\footnote{To simplify the illustration the \emph{compressed} public key
is not included in Figure~\ref{fig:overview}.}). Since any \emph{active} use
of crypto-assets generally requires a matching digital signature, the
underlying public key is necessarily revealed. This implies that the
\emph{same} cryptographic keying material, potentially used in
\emph{different} cryptocurrencies, can be linked to the same entity
controlling the respective key.

We analyzed six different cryptocurrencies: Bitcoin (BTC), Ethereum (ETH),
Litecoin (LTC), Dogecoin (DOGE), Zcash (ZEC) and Tron (TRX) and found that a
total number of \numReusedKeys\ keys have been reused in at least two of the
considered systems. Our results reveal that key reuse across cryptocurrencies
appears to happen at a considerable scale. Hereby, focusing on public key
reuse allows us not only to link entities across cryptocurrencies with
inconvertible address formats, but also to potentially link UTXO style
transactions with only one input. Moreover, in combination with other
(existing) clustering techniques, such as multiple-input heuristics, our
cross-cryptocurrency key reuse clustering method can also be used to cluster
account based designs such as Ethereum. Summarizing, we make the following
contributions:

\begin{itemize}
  \item We show that cryptographic keys are extensively and continually reused
    across all of the analyzed blockchain networks, with \numReusedKeys\ reused
    keys, of which at least \numActiveReusedKeys\ keys have been \emph{actively}
    reused.
  \item We present the first cross-chain key reuse analysis between UTXO and
    account models and show that a significant number of public keys are reused,
    namely \numReusedKeysBtcETH~between Bitcoin and Ethereum and
    \numReusedKeysUtxoAccount~in total between all of the analyzed UTXO and
    account models, accounting for roughly 50\% of all observed reused keys.
  \item We show that it is possible to cluster addresses associated with
    reused keys in account based cryptocurrencies based on their clustering in
    a UTXO cryptocurrency.
  \item To validate our approach, we reproduce the findings from related work
    that relies on address-based key-reuse
    detection~\cite{smolenkova2025keylinker,harrigan2018airdrops},
    and leverage similarities between Ethereum and Tron.
    Hereby, our key-based methodology can lead to further improvement by
    identifying reuse cases where direct address conversion between formats is
    not possible.
\end{itemize}

\section{Background}\label{sec:background}

Addresses are a core abstraction in cryptocurrency systems. They act as
pseudo\-nymous identifiers that bind public keys of digital signature schemes
to assets and thereby support publicly verifiable authorization of spends.
Reuse of an address (or its underlying key) weakens unlinkability and enables
entity inferences at scale, a phenomenon extensively leveraged in prior work
on tracing and clustering
\cite{nakamoto2008bitcoin,reid2012analysis,meiklejohn2013fistful}. In
UTXO-based designs, the \emph{multiple-input heuristic} was the first
widely-used clustering technique: inputs co-spent in a transaction are likely
controlled by the same entity
\cite{nakamoto2008bitcoin,reid2012analysis,meiklejohn2013fistful}. While
effective, it is
\begin{enumerate*}[label=\roman*)]
\item inapplicable to single-input models and
\item susceptible to intentional obfuscation (e.g., CoinJoin/PayJoin)~%
  \cite{glaeser2022foundations,ghesmati2021payjoin}.
\end{enumerate*}
The account model lacks a direct analogue to multiple-input heuristics. To
overcome this limitation Victor et al.~\cite{victor2020address} present
Ethereum-specific heuristics that exploit on-chain behavior, notably exchange
deposit address reuse, airdrop multi-participation, and token transfer
self-authorization, clustering a substantial fraction of active EOAs
(Externally Owned Accounts).

Depending on the system design, the \emph{same} public key may be associated
with different addresses within the same cryptocurrency, e.g. by using
different script types in Bitcoin-like UTXO models\footnote{See
Appendix~\ref{sec:utxo_types} for a listing of different Bitcoin script
types.}. While Kalodner et al.~\cite{kalodner2020blocksci} already briefly
mention this form of \emph{internal} key reuse\footnote{In \emph{BlockSci}
such addresses are referred to as \emph{equivalent},
see~\cite{blocksci-equiv-addresses}.}, its potential relevance, e.g. in the
context of clustering heuristics, is not discussed. In recent concurrent work,
Smolenkova and Yanovich address this issue and present a \emph{single chain}
address grouping method called \emph{KeyLinker}, that leverages convertible
address formats to identify internal key reuse in
Bitcoin~\cite{smolenkova2025keylinker}. A crucial difference to our work
however, aside from the single cryptocurrency setting, is that the presented
method relies on address conversion for reuse detection.
Their approach is therefore limited to convertible address formats and can not
detect key reuse in the case of incompatible address representations, such as
those derived from compressed and uncompressed public keys.

Cross-chain mechanisms (bridges, relays, atomic swaps, messaging protocols)
now underpin asset mobility and composability across heterogeneous ledgers
\cite{zamyatin2021sok} and play an important role in decentralized finance
applications. In parallel, empirical work has begun to measure bridge behavior
and risks (costs, inconsistencies, attack surfaces), underscoring the need for
transparent, on-chain-verifiable linkage
\cite{augusto2024sok,yan2025empirical,lin2025connector}. Prior cross-ledger
studies often target \emph{transaction} association rather than \emph{entity}
resolution. Yousaf et al. combine service (ShapeShift) data with timing/amount
constraints to link flows across chains \cite{yousaf2019tracing}. CLTracer
generalizes cross-ledger tracing using address-relationship heuristics
\cite{zhang2022cltracer}. For \emph{contract}-based bridges, CONNECTOR
replaces API/ledger dependence with purely on-chain semantics to match
deposits and withdrawals \cite{lin2025connector}. Recent measurement work
further analyzes cost asymmetries, ledger inconsistencies, and other
characteristics of (EVM-compatible) cross-chain transactions
\cite{yan2025empirical}.

Closer to our focus on cross-chain \emph{public key reuse}, Harrigan et al.\
(HSI) demonstrate that an airdrop in CLAM, which reused address encodings
from other UTXO chains, enables \emph{cross-chain address clustering} across
BTC, LTC and DOGE and can reveal ownership on one chain through activity on
another \cite{harrigan2018airdrops}. BlockSci also provides a
\emph{Multi-chain} mode for Bitcoin-like UTXO models that allows for address
based cross-chain clustering similar to HSI. The paper also analyzes key-reuse
between Bitcoin and the \emph{Bitcoin Cash} hard
fork~\cite{kalodner2020blocksci}. In \cite{hinteregger2019short} the impact of
hard forks on cross-chain traceability in Monero is analyzed by leveraging the
reuse of mixins across forks.

Orthogonal to service- or transaction-layer analysis, others examine public
key reuse for different purposes. Breitner and Heninger collect and analyze
keys across cryptocurrencies to expose biased or repeated ECDSA nonces
\cite{breitner2019biased}. Li et al.\ show deanonymization risks from
public-key reuse at Ethereum’s \emph{P2P layer} (networking level), distinct
from transaction-layer reuse exploited here \cite{li2025egninfoleaker}.
Industry has also proposed patented methods to correlate identities across
chains via public-key–derived hashes on common curves \cite{US11438175B2}.

These previous lines of work motivate and contextualize our approach: while
prior efforts either presume identical address formats (EVM) or leverage
shared or convertible encodings among UTXO-like systems
\cite{harrigan2018airdrops,kalodner2020blocksci,smolenkova2025keylinker}, we
exploit \emph{cross-chain public-key reuse} to bridge the heterogeneous
designs of the UTXO \emph{and} account model, thereby expanding the space of
cross-chain/entity linkage beyond address-string equality.

\section{Dataset and methods}\label{sec:data_methods}

Most cryptocurrency transaction types do not directly disclose the public key
of a recipient. Mainly due to efficiency and space reasons an encoded
cryptographic hash of the public key, commonly referred to as an
\emph{address}, is used instead of the raw public key~\cite{bonneau2015sok}.
The underlying key is revealed when funds are first spent, since producing a
spending transaction requires a signature that exposes the associated key for
validation. Therefore, almost all public keys we collected for our analysis
have been \emph{actively} used on the sending side at least once, e.g.,
because they were used to sign a cryptocurrency transaction.

We gathered transaction data from six representative cryptocurrencies (BTC,
LTC, DOGE, ZEC, ETH, and TRX) ending with the blocks mined just prior to
April 1, 2025 (cf. Table~\ref{tab:blockchain-stats} in
Appendix~\ref{sec:block_ranges}). Note that none of the considered systems
shares a common transaction history, i.e. none is a permanent fork of
another~\cite{zamyatin2018wild}, which could skew our analysis of key reuse.
These transactions have been used to extract \emph{secp256k1} public keys.
Since retrieving public key information is not immediately possible through
common APIs, the extraction procedure had to be adapted depending on the
cryptocurrency as well as the respective transaction type. It is
generally\footnote{Exceptions such as Bitcoin \emph{Pay to Public Key (P2PK)}
or \emph{Pay to Multisig} (P2MS) transactions can exist.} not possible to
extract the public key directly from addresses without a spending transaction,
due to the preimage resistance properties of the underlying cryptographic hash
function~\cite{katz2020modern} predominantly used in address generation.
Passive usage, where funds are only ever received at an address, is also
the reason why we distinguish between two different classes of key reuse.

\subsection{Classification of Key Reuse}\label{subsec:classification_key_reuse}

For our analysis, we differentiate between \emph{active-} and \emph{passive
key reuse} (see Definition~\ref{def:key-reuse}). Active key reuse indicates
the usage of the same secret key for signing (i.e., sending) a transaction in
more than one cryptocurrency, while passive key reuse indicates that the same
address has been used on the receiving side of a transaction in more than one
cryptocurrency.

\begin{definition}\label{def:key-reuse}[Cross-chain Key Reuse]
If asymmetric key pairs are used in more than one cryptocurrency in at least
one transaction each, either on the sending or receiving side, we call this a
(cross-chain) \emph{key reuse event}, or a \emph{reused key} respectively. We
further differentiate between:
\begin{itemize}
  \item \textbf{active key reuse}: When the keying material is actively used
    to sign e.g., a transaction, in at least two cryptocurrencies, we call
    it an \emph{active} key reuse.
  \item \textbf{passive key reuse}: When the keying material is actively used
    for signing in at most one cryptocurrency and for all other
    cryptocurrencies the key is only passively involved on the receiving side,
    we call it a \emph{passive} key reuse.
\end{itemize}
\end{definition}

We introduce this separation since passive key reuse is not preventable by the
owner of the respective key. Specifically, once the public key has been
disclosed anywhere or the associated address(es) can otherwise be inferred,
transactions that are directed towards these addresses can generally not be
prevented. Active reuse, on the other hand, requires knowledge of the
respective private key. Therefore, it is a strong indicator that the same
entity is involved in those transactions\footnote{Exceptions to this rule
would be key loss/compromise or the use of threshold
ECDSA~\cite{lindell2018fast,wong2023real}, which would allow multiple
potential signers to produce a single signature. At the time of writing this
is however not widely used.}. In the six cryptocurrencies that we analyzed,
\numReusedKeys\ keys have been reused in at least two of them, of which
\numActiveReusedKeys\ keys have been \emph{actively} reused in more than one
cryptocurrency.

\subsection{Data Extraction}\label{subsec:data_extraction}

Our proposed analysis technique for key reuse requires that the respective
public keys have been revealed at least once. However, most cryptocurrency
transaction types do not immediately disclose the public key of a recipient.
The only exceptions to this rule in our dataset are \emph{Pay to
Public Key} (P2PK)~\cite{learnmeabitcoin_p2pk} and \emph{Bare Multisig}
(P2MS)~\cite{learnmeabitcoin_p2ms} transactions in Bitcoin and Litecoin.
The use of P2PK is predominantly documented in the coinbase
transactions of the blockchains' initial blocks. The P2MS script type is a
historical artifact within the Bitcoin protocol, rendered largely obsolete by
the introduction of \emph{Pay to Scripthash} (P2SH) in 2012. Both transaction
types have long been deprecated and account for only a small number of
transactions (e.g. for Bitcoin \numPtoPK\ P2PK (0.11\%) and \numPtoMS~P2MS
(0.12\%) transactions, respectively). 

\subsubsection{UTXO Cryptocurrencies}\label{subsubsec:utxo_currencies}

A modified version of \emph{bitcoin-etl}~\cite{bitcoin-etl} (version 1.5.2)
was used to extract and store transaction data for all considered UTXO
cryptocurrencies in JSON format. This modified version included additional
fields to capture Segregated Witness (SegWit) data, which is essential for a
comprehensive analysis of UTXO-based cryptocurrencies. UTXO outputs can be
created with different types of locking scripts, which define the conditions
under which those specific coins can be spent in the future (see
Appendix~\ref{sec:utxo_types} for details). Over time, UTXO output types have
evolved to improve efficiency, privacy, and functionality~\cite{jain2023sok}.
While these scripts often correspond to standard addresses (e.g., P2PKH or
P2WPKH), they can also represent more complex conditions such as
multisignature requirements or time locks. A single transaction might create
outputs with different spending conditions for various purposes.

To extract the ECDSA public keys associated with a respective address, we had
to adapt the extraction to the particular transaction type (cf.
Appendix~\ref{sec:utxo_extract}). Table~\ref{tab:num-extracted-pubkeys} shows
the total counts of extracted unique public keys, split into the respective
output types.

\begin{table*}[ht]
  \centering
  \resizebox{0.975\textwidth}{!}{\begin{tabular}{lrrrrrrr}
  \toprule
Currency & P2PK & P2MS & P2PKH & P2SH & P2WPKH & P2WSH & Total count \\ 
  \midrule
BTC & 217,060 & 3,543,439 & 580,334,145 & 517,671,478 & 302,098,052 & 59,772,637 & 1,462,721,776 \\ 
  LTC & 300,688 & 34 & 65,420,695 & 164,912,000 & 121,245,406 & 1,144,188 & 352,973,720 \\ 
  DOGE &  &  & 82,459,658 & 1,873,438 &  &  & 84,332,852 \\ 
  ZEC &  &  & 7,434,678 & 1,125,355 &  &  & 8,560,024 \\ 
   \bottomrule
\end{tabular}
}
  \caption{Number of UTXO transaction types and total counts of extracted
    unique public keys.}
  \label{tab:num-extracted-pubkeys}
\end{table*}

\subsubsection{Account Based Cryptocurrencies}\label{subsubsec:account_based_currencies}

For our analysis we parsed all Ethereum transactions initiated by EOAs and
extracted a total of \numUniqueEth\ unique public keys using the public key
recovery method described in Appendix~\ref{sec:account_based_extract}. This
was necessary because the Ethereum node that was used to query blockchain data
(Geth~\cite{geth}) does not provide a way for retrieving the public key
directly. The code we developed for this recovery is provided on
GitHub\footnote{\url{https://gist.github.com/kernoelpanic/423c61f90e81e4c9d473ff6fda783559}}.
Due to the large number of Tron transactions (cf.\ %
Table~\ref{tab:blockchain-stats} in Appendix~\ref{sec:block_ranges}), we did
not perform signature recovery for every transaction and instead first
identified the relevant subset of \numUniqueTrx\ transactions where a new
source/owner address (i.e. EOA) first performed a transaction before
recovering the unique associated public keys. Furthermore, we restrict our
extraction of public keys in ETH and TRX to EOAs and leave other scenarios,
such as signature verification in smart contract executions, to future work.

\subsection{Validation of our Approach}\label{subsec:validation}

To validate our approach of identifying cross-chain key reuse through the
underlying public keys, we compare our results with:
\begin{enumerate*}[label=\roman*)]
\item The results presented in HSI~\cite{harrigan2018airdrops} on address
  reuse in Bitcoin, Litecoin and Dogecoin
\item A ground truth between Ethereum and Tron which we are able to obtain due
  to their mostly compatible address format and
\item \emph{Internal}, i.e., single chain key-reuse in Bitcoin with the
  results in~\cite{smolenkova2025keylinker}, which we discuss in
  Subsection~\ref{subsubsec:utxo_internal_reuse}.
\end{enumerate*}

The approach in HSI relies on the fact that all of their considered systems
use address formats that encode a hash digest of the respective public key or
Bitcoin-compatible spend script derived by the same hash function. Therefore,
these hashes are used for comparison (not the entire address). The advantage
of this approach: It can also identify reuse of addresses (and thus keys)
which are used \emph{exclusively passive}. The disadvantage is that this
technique only works across cryptocurrencies with the same underlying address
hash formats.

By computing our public key based analysis for the block range in HSI we are
able to compare our approach against their results\footnote{See Appendix
Section~\ref{subsec:bitcoin_litecoin_doge_validation} for details.}.
As expected, when relying only on public keys the instances where
\emph{exclusively passive} cross-chain reuse occurs cannot be identified.
We identified \numPercentHarrignBtcLtc\% of the HSI reuse cases for
Bitcoin-Litecoin, \numPercentHarrignBtcDoge\% of cases for Dogecoin-Bitcoin,
and \numPercentHarrignDogeLtc\% for Dogecoin-Litecoin. Surprisingly,
when comparing our methodology to HSI's up to our current block range the
public key based approach exceeds expectations, ranging from
\numPercentHarrignAllLtcBtc\% for Bitcoin-Litecoin to
\numPercentHarrignAllBtcDoge\% for Bitcoin-Dogecoin and
\numPercentHarrignAllLtcDoge\% for Litecoin-Dogecoin. Possible explanations
are that HSI's method does not work between P2WSH (SHA256) and P2SH (RIPEMD160),
as well as compressed and uncompressed keys. Moreover, Dogecoin makes extensive
use of P2SH, for which address based reuse checks are less effective compared
to public key extraction, e.g., in case of multi-sig transactions.

For validating our approach in account based systems, we rely on the fact that
Ethereum and Tron share the same underlying Ethereum address format, where
Tron just encodes a prefixed version of the Ethereum address with
Base58\footnote{See Appendix Section~\ref{sec:ethereum_tron_validation} for
details.}. Decoding Tron addresses to bring them into the same format allows
us to obtain a ground truth which also captures addresses which are used
\emph{exclusively passive}. Compared to this ground truth, we were able to
identify \numTrxEthActivePercent\% of reuse events based on our method of
public key extraction, which is in line with our previous comparisons,
suggesting that around 90\% of cross-chain key reuse can be identified when
exclusively relying on public keys. This provides a strong indicator for the
validity of the approach also in instances where no ground truth is available
and address formats are not readily convertible, as in the case of reuse
between UTXO and account based systems.

We note that due to the public nature of the cryptocurrencies that we
analyzed, all transaction data is available, allowing our results to be
reproduced.

\section{Analysis \& Results}\label{sec:analysis_results}

\subsection{Active Public Key Reuse across Different Blockchain Networks}\label{subsec:key_reuse_crosschain}

\begin{figure*}[ht]
  \centering
  \includegraphics[width=1\textwidth]{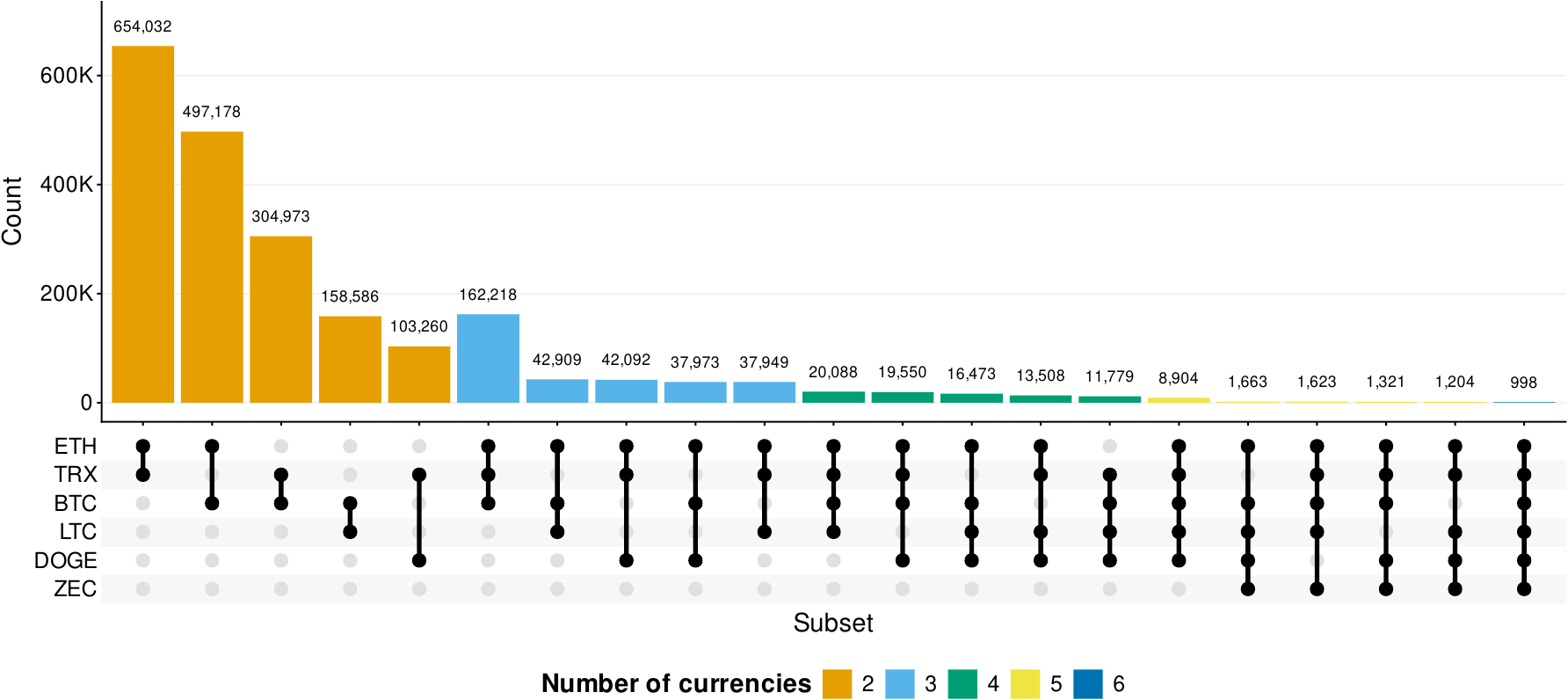}
  \caption{Cardinality of set intersections for active public key reuse across
    the considered currencies (Top 5 cardinalities per intersection size).}
  \label{fig:upset-plot}
\end{figure*}

Building on the extracted dataset, we analyze active cross-chain public key
reuse. By treating keys from each cryptocurrency as discrete sets, we identify
cross-network reuse through set intersections. We visualize these
intersections using an UpSet plot, which offers superior scalability over
traditional Venn diagrams for multi-set comparisons~\cite{lex2014upset}. The
plot uses bar charts to denote cardinality and a matrix layout to indicate
specific set memberships; Figure~\ref{fig:upset-plot} highlights the Top 5
cardinalities for each intersection size. For a complete representation of all
identified intersections we refer to Appendix~\ref{sec:appendix_active_reuse},
Figure~\ref{fig:upset-plot-full}.

We identified a total of \numReusedKeys\ reused keys. Of these, at least
\numActiveReusedKeys\ ~public keys were actively used for sending transactions
across more than one cryptocurrency. This distinction is crucial, as it
indicates a deliberate action by the key holder. This finding suggests a
notable level of cross-chain user activity, which can be further investigated
to understand user behavior, wallet design, and potential security and privacy
implications.

\subsection{Temporal Analysis}\label{subsec:temporal_analysis}

Whether public key reuse is a phenomenon of the past or present is a key
question. While early Bitcoin users might have reused keys due to a lack of
privacy awareness and immature wallet software, it is unclear if this behavior
has persisted. Given that UTXO-based cryptocurrencies permit multiple address
formats to be derived from a single key, our analysis proceeded in two stages:
First, we examined key reuse within single UTXO chains. Second, we conducted a
longitudinal investigation into public key reuse across distinct blockchain
networks.

\subsubsection{UTXO Internal Active Reuse}\label{subsubsec:utxo_internal_reuse}

\begin{figure*}[htbp]
  \centering
  \includegraphics[width=0.95\textwidth]{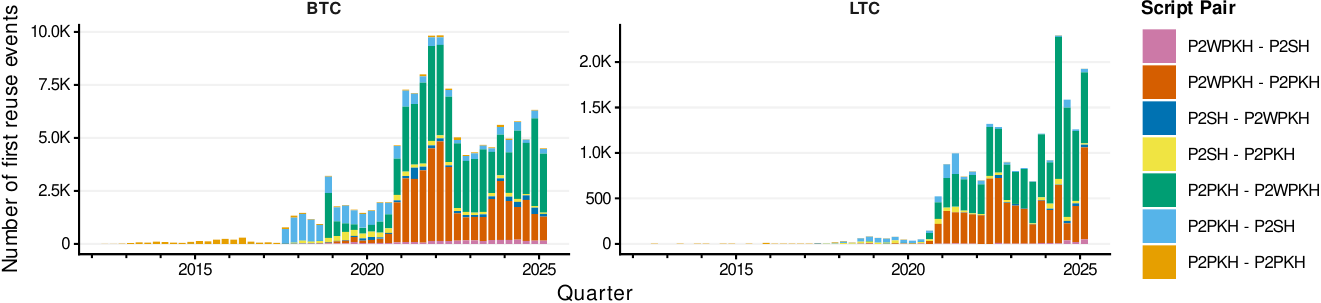}
  \caption{Quarterly counts of first UTXO internal active public key reuse
    events within Bitcoin (BTC) and Litecoin (LTC).}
  \label{fig:utxo-internal-reuse}
\end{figure*}

Figure~\ref{fig:utxo-internal-reuse} shows the quarterly counts of UTXO
internal reuse events for Bitcoin and Litecoin, revealing that internal public
key reuse on these networks remains a current phenomenon. Hereby, our results
confirm the findings in~\cite{smolenkova2025keylinker} that relies on
convertible address formats to detect internal key-reuse in Bitcoin. We
further show that while internal reuse can also be observed in other UTXO
cryptocurrencies, e.g. Litecoin, the occurrences may be very low (Zcash). The
ordering of the address formats highlights which format was first used for a
public key. A significant concentration of UTXO internal active reuse events
is observed within two specific public key type pairs: The pair P2PKH-P2WPKH
accounts for the largest proportion at 44.6\%, while the converse order
P2WPKH-P2PKH follows with 32.9\%. Combined, these two pairs represent
approximately 77.5\% of the total internal reuse events, indicating their
dominant role in the observed reuse patterns. The remaining public key type
pairs account for the residual proportion of 22.5\%.

Note that P2PKH-P2PKH internal reuse (1.8\%) refers to instances where one
address was derived from the uncompressed format of the public key and the
other from the compressed format. This occurs in either order, meaning the
first transaction could use an address from an uncompressed key and the
subsequent transaction an address from a compressed key, or vice versa.

The internal reuse analysis of Zcash revealed a total of four instances of
internal public key reuse. Conversely, in Dogecoin a median of two first
active internal key reuse events were documented per quarter, peaking at
168~events within the first quarter of 2018.

\subsubsection{Cross-chain Active Reuse}\label{subsubsec:crosschain_reuse}

\begin{figure*}[htbp]
  \centering
  \includegraphics[width=0.95\textwidth]{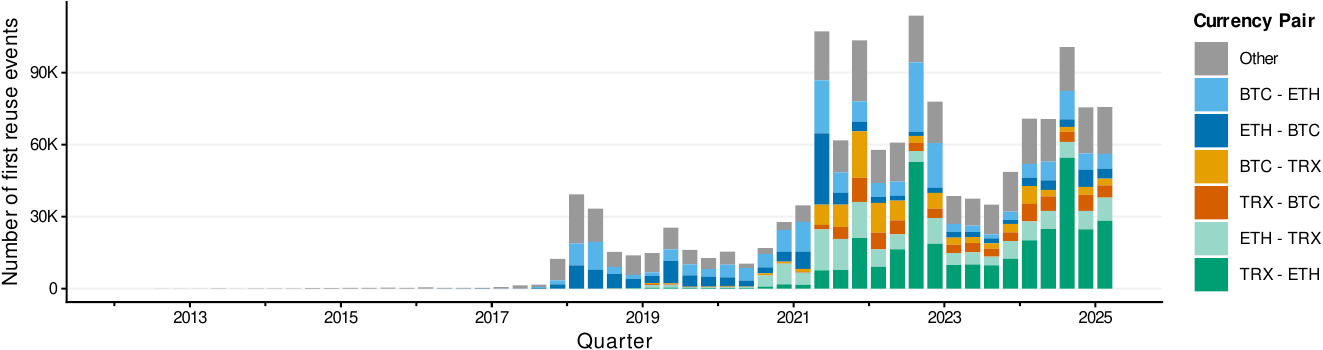}
  \caption{Quarterly counts of first active public key cross-chain reuse events.}
  \label{fig:crosschain-reuse}
\end{figure*}

The cross-chain reuse of public keys over-time is shown in
Figure~\ref{fig:crosschain-reuse}. Again, the order of the cryptocurrency
tickers indicates in which cryptocurrency the underlying public key was used
first. The six most frequent currency pairs involve the three dominant
cryptocurrencies BTC, ETH, and TRX. Their reuse events constitute
approximately 73.3\% of all observed first active cross-chain reuse events,
highlighting their central role in the observed transactions. Notably, the
TRX-ETH pair accounts for the largest proportion of these events, at 23.3\%.
Here, we observe that it is more common that public keys are first used in TRX
before they are used in ETH, than the other way round. The remaining currency
pairs not listed (category \emph{Other} in Figure~\ref{fig:crosschain-reuse})
collectively represent the remaining 26.7\% of reuse events, indicating a
significant concentration of activity within the six most frequent pairs.

\subsection{Improve any Clustering through Reused Keys}\label{subsec:clustering}

Since the (re)use of the same cryptographic key necessitates control over the
associated private key, this is a strong indicator that the same entity is
behind all derived cryptocurrency address formats across the respective
chains. This rationale can be used to improve any existing clustering method
by merging clusters with different addresses derived from the same underlying
public key.

Depending on the cryptocurrency design, it may be possible to associate public
keys with more than one address, in which case this rationale can also be
applied to a \emph{single}
cryptocurrency\footnote{Figure~\ref{fig:cluster_int} in the Appendix shows the
basic process on the example of Bitcoin.}. In concurrent work Smolenkova and
Yanovich introduce and discuss such a single chain address clustering method in
the context of Bitcoin~\cite{smolenkova2025keylinker}. Their methodology,
however, relies on convertible address formats rather than extraction of the
underlying public keys.

The primary goal of our work is to analyze and quantify \emph{cross-chain} key
reuse, whereby address formats may not be directly convertible and hence
extraction and reliance on the underlying public keys is necessary. By
first applying our key-based approach to the single chain setting, we are not
only able to verify the results from~\cite{smolenkova2025keylinker}, but can
also provide a lower bound on the internal key reuse events that cannot be
captured by address conversions. Interestingly, we find that for internal key
reuse across the analyzed UTXO designs the reuse between compressed and
uncompressed public key formats only accounts for 1.8\% of all observed
key-reuse events and 1.99\% in the case of Bitcoin only.

Our analysis method starts with the initial clustering derived from the
multi-input heuristic. In a second step, we merge all clusters (i.e.,
recompute the connected components) which contain addresses derived from the
same public key. Using this technique we were able to merge 2,040,417
clusters, reducing the overall number of clusters from 618,812,655 to
616,772,238.

\subsection{A Cross-Chain Clustering Method for Account-Based Systems}\label{subsec:clustering_account}

By relying only on actively reused keys, the derived addresses in
cryptocurrencies with different designs can be associated with a common entity
controlling the respective private key. A new non-heuristic based clustering
method is to use a retrieved public key in \emph{UTXO based} cryptocurrency
$A$ to derive potential addresses in \emph{account based} cryptocurrency $B$,
thereby associating those addresses to the same entity across different system
designs. This also works if addresses (and their associated keys) in
cryptocurrency $B$ have not yet been actively used in $B$, but the same
underlying key has been actively used in $A$.

\begin{figure}[ht!]
  \centering
  \begin{minipage}{0.45\textwidth}
    \includegraphics[width=\linewidth]{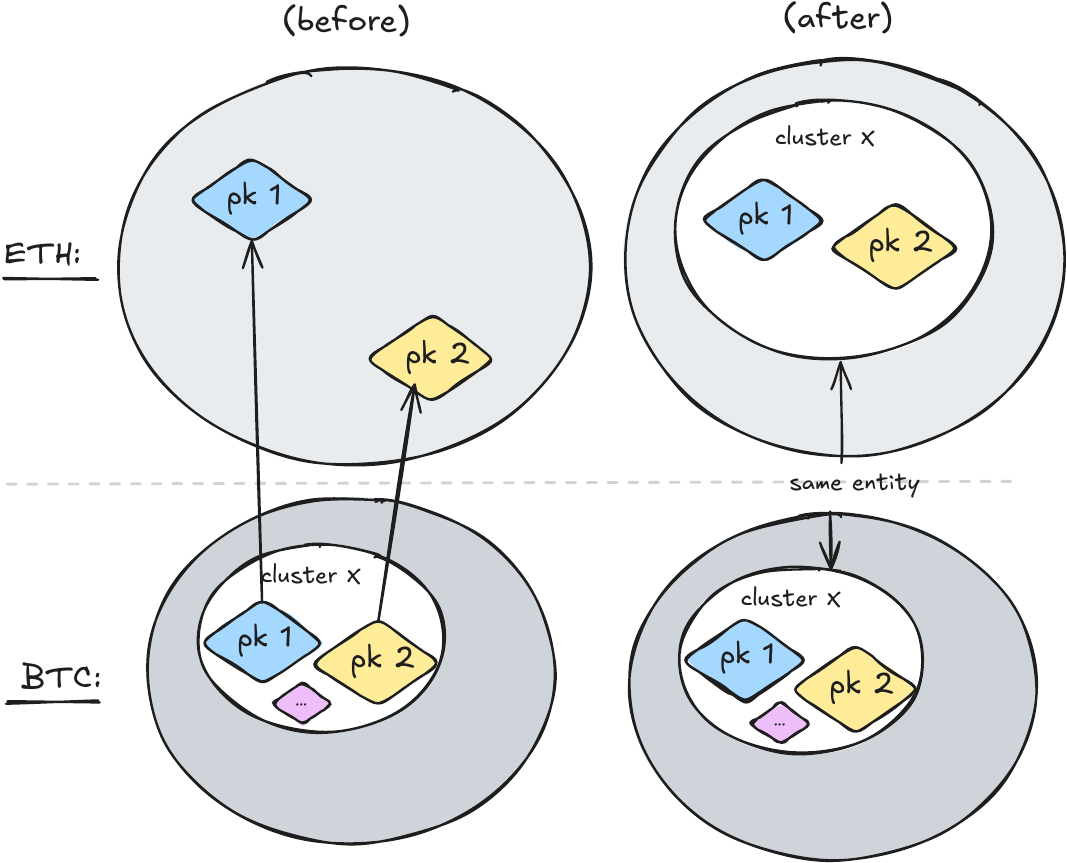}
  \end{minipage}%
  \hfill
  \begin{minipage}{0.475\textwidth}
    \caption{Example of clustering an account-based cryptocurrency such
      as Ethereum, based on multiple-input clustering in Bitcoin.}
    \label{fig:cluster_acc}
  \end{minipage}
\end{figure}

Building on this basic method, any heuristic based clustering in
cryptocurrency $A$ can be used to also cluster addresses in another
cryptocurrency $B$, for which no clustering is readily available yet, e.g.,
because $B$ is account based while $A$ is UTXO based, which enables the
application of the multiple-input heuristic. Any available clustering in $A$
can be transferred to $B$ by computing all derivable addresses (applicable to
currency $B$) from obtained public keys within a single cluster in $A$ and
assigning them to a new cluster in $B$, if the respective addresses already
exist there. Thereby, multiple-input heuristic clustering from UTXO based
cryptocurrencies can be used to cluster account based cryptocurrencies.
Figure~\ref{fig:cluster_acc} shows an illustration using Ethereum and Bitcoin
as example. Using our techniques we were able to cluster 497,289 Ethereum
addresses into 62,681 clusters, and 305,019 Tron addresses into 19,242
clusters, respectively.

\subsection{Entities performing Key Reuse}\label{subsec:entities_reuse}

To determine whether key reuse is more prevalent among services and companies
or individual end-users, we analyzed attribution data associated with
addresses of reused public keys. If these addresses are predominantly linked
to known services, it would suggest that services are responsible for the
majority of key reuse incidents. Conversely, a lack of such service-specific
tags would indicate that end-users are the primary contributors. This approach
allows us to empirically analyze the distribution of responsibility for this
phenomenon. By merging the addresses exhibiting public key reuse with
attribution data from Iknaio~\cite{iknaio-platform}, which builds on the
open-source analytics platform GraphSense~\cite{haslhofer2021graphsense}, we
identified some entities responsible for key reuse.

A key finding of this analysis is the direct attribution of 1,200 addresses
(originating from 894 unique instances of key reuse) to two prominent
cryptocurrency exchanges. According to CoinMarketCap
rankings~\cite{coinmarketcap}, both entities are among the Top 10 spot
exchanges by volume. Specifically, these entities accounted for 792 and 102
reused public keys, respectively. This linkage indicates the contribution of
major services to the observed instances of key reuse. While pairwise overlaps
are common, we also observe higher-order key intersections involving three or
more distinct blockchains (ca.\ 26\%) in these exchange-linked instances. This
distribution suggests a systemic reliance on multi-currency key reuse within
the operational frameworks of some top-tier spot exchanges.

By cross-referencing public key data with address labels sourced from Dune
Analytics~\cite{dune-analytics} and Iknaio, we also identified significant key
reuse across several other functional categories. Specifically, our analysis
revealed 1,297 reused keys associated with \emph{Name Services}, followed by
\emph{Contract Deployers} (1,070), \emph{DeFi Bridges} (911), and \emph{Mixing
services} (203)\footnote{We conduct a manual investigation of the key reuses
in mixing services in Appendix~\ref{sec:tornadocash}}.

\begin{figure}[ht!]
  \centering
  \begin{minipage}{0.5\textwidth}
    \includegraphics[width=\linewidth]{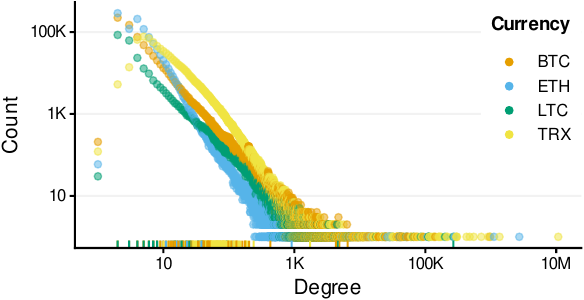}
  \end{minipage}%
  \hfill
  \begin{minipage}{0.45\textwidth}
    \caption{Degree ($x$-axis) distribution in address networks for addresses
      corresponding to reused public keys. The $y$-axis is the count of nodes
      having the respective degree. The ticks above the $x$-axis correspond to
      the degrees of specific addresses that have been identified as belonging
      to exchanges.}
    \label{fig:degree-distribution}
  \end{minipage}%
\end{figure}

To further understand the network characteristics of key reuse, we constructed
a network where nodes represent addresses and undirected edges represent
aggregated transactions (incoming or outgoing). As a measure of centrality, we
then computed the degree for each address, representing the total number of
neighbors with which this address has interacted.
Figure~\ref{fig:degree-distribution} illustrates the degree distribution for
four cryptocurrencies under consideration. We also mark degrees of nodes with
addresses that correspond to attribution data for known exchanges with a small
tick above the $x$-axis. An interesting finding in this regard is the presence
of these attributed exchange addresses also across the lower degree spectrum,
indicating the use of per-customer addresses, which might then get aggregated.
Addresses with a high degree, represented by points in the lower-right (above
degree 1,000) of the figure, are potentially associated with services or
traders which manage a large volume of transactions. This, as well as the
attribution of certain addresses to exchanges, indicates that key reuse is not
just happening in private, but also professional contexts.

\section{Discussion and Future Work}\label{sec:discussion}

For the analyzed cryptocurrencies we dismiss several explanations as the main
cause for the observed cross-chain key reuse. An intuitive reason could be the
existence of a prior hard fork or other forking event that led to a permanent
chain split~\cite{zamyatin2018wild,kalodner2020blocksci}. However, none of the
cryptocurrencies herein considered share such a common transaction history
with each other.

While we identified 279 weak cryptographic private keys (i.e., private key is
a low number $<2^{32}$ or a $256$ bit string of hamming weight $<5$) in the
set of reused keys, the fact that a significant portion continue to hold value
as of today (e.g., 88\% of ETH keys hold value and collectively control more
than 1 million Ether in total) renders an explanation involving weakly
generated cryptographic keys highly unlikely.

Cross-chain \emph{airdrops} could, in principle, offer another explanation but
we are not aware of any major events that would help explain the observed
scale of key reuse. In HSI the airdrop was performed on the Clams network,
which is not part of our analysis. For Tron, we considered a major airdrop and
token migration when the project launched its own independent Blockchain,
however it appears that the addresses associated with the ERC20 token on
Ethereum were not automatically reused. Further, unlike most EVM-based chains
Tron is not directly compatible with MetaMask and defines its own SLIP44
coin\_type \cite{rusnak2014slip44}, leading to a different key derivation
path.

Neither of the previous examples can readily offer an explanation why key
reuse also occurs across UTXO and account-based designs, which is the main
focus of our analysis. Cross-chain transfer
technologies~\cite{zamyatin2021sok,augusto2024sok} may yield a plausible
answer, in particular since prior work has outlined that address reuse in
bridges between EVM-compatible designs appears to be a common
occurrence~\cite{lin2025connector,yan2025empirical}. However, this observation
is only made for cross-chain transfers in account-based cryptocurrencies with
\emph{exactly the same} address format. Specifically, for EVM-compatible
designs the associated wallet software often actively encourages key reuse,
as we will discuss below.

We are not aware of any technical requirement for cross-chain transfer
mechanisms or bridge designs between UTXO and account-based models that
necessitate reusing the same secp256k1 keys in both systems. However, it does
appear that there exist bridge designs which intentionally perform key reuse
between UTXO and account models to improve usability, namely the trusted
execution based Avalanche Bridge\footnote{A detailed discussion can be found
in Appendix~\ref{sec:tornadocash}.}. Note that while this bridge primarily
operates in the Avalanche cryptocurrency, which we do not consider in this
paper, there exists an observable form of \emph{spill-over} in cases where
further key reuse occurs between Avalanche and Ethereum. Although we were able
to attribute \emph{DeFi Bridges} as one of the functional categories
associated with reused keys (see Section~\ref{subsec:entities_reuse}), our
results do not suggest that they are responsible for the bulk of the
significant number of identified reuse events between UTXO and account based
models.

Similarly, while our analysis regarding the entities performing key reuse has
revealed that large exchanges are also among those entities, the underlying
question why keys are reused between UTXO and account based systems cannot
conclusively be answered yet.

\subsubsection{Deterministic Key Derivation as the Culprit?}\label{subsubsec:deterministic_key_derivation}

According to today's best practices, keys in cryptocurrency systems should be
deterministically derived and ideally only used once for a single transaction.
Such \emph{hierarchical deterministic (HD) wallets} were standardized and
subsequently expanded upon in e.g., BIP32~\cite{wuille2012bip32},
BIP39~\cite{palatinus2013bip39}, BIP44, etc., in order to replace earlier
solutions, which generally relied on randomly (pre)generated keys stored in a
keyfile. These deterministic standards are now ubiquitously implemented across
diverse cryptocurrency networks and hardware/software wallet
providers~\cite{ledger,trezor}.

It is conceivable that deterministic key derivation is partially responsible
for cross-chain key reuse~\cite{kalodner2020blocksci}. While standards such as
BIP44 aim to provide sufficient domain separation between cryptocurrencies by
defining derivation hierarchies that specify different \emph{coin\_types},
adherence to this pattern is not guaranteed.
For example, popular wallet software for EVM-compatible networks, such as
MetaMask~\cite{metamask} and Exodus~\cite{exodus-derivation-paths},
\emph{actively encourages key reuse} by deriving keys from a single
\emph{coin\_type} derivation branch for different systems and allowing users
to seamlessly ``switch networks'', i.e. different cryptocurrencies, while
retaining the same keys/accounts~\cite{metamask2025ethereum}. From a usability
and convenience perspective such features may prove themselves helpful, e.g.
by allowing the user to more readily memorize and recognize their personal
account address(es) across different networks. Unfortunately, this practice
not only reduces privacy, but also security, since compromising the underlying
private key would also lead to loss of funds in multiple systems. It can also
help explain results from previous research that observes address reuse in
EVM-compatible designs while analyzing and tracking cross-chain
transactions~\cite{lin2025connector,yan2025tracing}.

The accidental or intentional reuse of the same derivation branch in HD
wallets for different cryptocurrencies is likely not just limited to
EVM-compatible designs. It therefore stands to reason that our dataset may
have also captured HD wallet reuse between UTXO and account-based designs.
However, a reliable verification of this hypothesis does not appear to be
straightforward\footnote{We discuss a possible weak indicator of HD wallet
reuse in Appendix~\ref{app:hd-method}.} and we leave a detailed investigation
to future work.

\subsection{Implications}\label{subsec:implications}

Our observation of significant and active key reuse across different
cryptocurrencies and system designs has several implications. Cross-chain key
reuse occurs even across fundamentally different system designs (UTXO and
account-based), yet our analysis cannot conclusively determine whether this
practice is intentional or inadvertent; this highlights the need to better
understand its causes and to raise awareness that such reuse can weaken both
security and privacy.

From a privacy and forensic viewpoint, key-based cross-chain attribution
provides ground-truth entity links that do not rely on heuristic clustering
rules and can thus be used both to evaluate existing heuristics and to extend
UTXO-style clustering techniques into account-based systems. In our entity
analysis we also attribute 203 reused public keys to mixing services. Such
behavior creates cross-chain identifiers that may enable deanonymization and
can complement existing clustering and tracing approaches, even in contexts
explicitly designed to obfuscate transaction flows.

\subsection{Countermeasures}\label{subsec:countermeasures}

From a privacy and security point of view, key reuse is undesirable. Therefore
the prime objective should be to prevent it e.g., by ensuring \emph{domain
separation} in deterministic key derivation and implementing appropriate usage
patterns that discourage key reuse. There are plenty of reasons for deriving,
ideally short-lived, cryptographic keying material. Firstly, the usage of
per-transaction keys in UTXO based cryptocurrencies makes linking transactions
of the same entity harder, increasing the privacy of the user. It has
therefore been recommended as best practice right from the inception of
Bitcoin~\cite{nakamoto2008bitcoin}. Moreover, the usage of per-cryptocurrency
derived keys also prevents certain security related issues, such as the replay
attacks which happened during the Ethereum Classic hard
fork~\cite{Buterin2016EIP155}, or other attacks such as accidental reuse, or
overlapping nonces~\cite{breitner2019biased} in context of ECDSA. Last but not
least, mitigating risks by \emph{not putting all your eggs in one basket} is
generally considered good practice.

\subsection{Limitations}\label{subsec:limitations}
\begin{itemize}
  \item For UTXO based multi-signature transactions, we did not attempt to
    identify the actively used public keys for signatures in a $t$ out of $n$
    multi-signature transaction. Therefore, we did not consider any public keys
    retrieved from P2MS and P2SH multi-signature transactions as
    \emph{actively} used.
  \item For Zcash, we restricted the analysis to the transparent pool
    (t-pool), which is comparable to Bitcoin regarding its functionality.
    The shielded pool (which utilizes zk-SNARKs) does not provide the required
    transaction details.
  \item We limited our analysis of Bitcoin and Litecoin to transaction types
    using legacy as well as SegWit protocols. We did not consider transactions
    that utilize more recent protocol upgrades, specifically
    Taproot~\cite{jain2023sok}, for both Bitcoin and Litecoin, and the
    \emph{Mimblewimble Extension Blocks} (MWEB) on Litecoin. This decision was
    made to ensure dataset consistency and to avoid
    complexities introduced by the newer, non-standard transaction formats.
  \item We limited our analysis of account based systems to EOAs, leaving the
    analysis of smart contract input data to future work.
\end{itemize}

\section{Conclusion}\label{sec:conclusion}

We show that more than 1.4~million secp256k1 keys have \emph{actively} been
reused within and across cryptocurrencies, and that this is not a phenomenon
of the past. Our measurements characterize key reuse both within individual
cryptocurrencies and across fundamentally different designs and reveal that a
wide range of entities such as exchanges, bridges, and mixers reuse keys. By
relying directly on public keys we provide ground-truth relationships that
span multiple systems and are independent of heuristic clustering rules.
Further, we demonstrate that key reuse enables the application of clustering
techniques that may only be available in certain designs to otherwise
incompatible systems, thereby improving the systematic identification of
cross-chain entity behavior.

\section{Acknowledgments}\label{sec:acknowledgments}

This research was partially funded by the Austrian KIRAS program (BMF) under
projects DeFiTrace (905300) and LLEA (926183), the FFG Bridge project SecKey
(46322124), the SBA Research COMET Center (SBA-K1 NGC), the University of
Vienna, and the European Union's Horizon Europe programme (ENSEMBLE,
grant agreement No.~101168360).

\printbibliography

\clearpage

\appendix

\section{Analyzed cryptocurrencies and block ranges}\label{sec:block_ranges}

\begin{table*}[ht]
  \centering
  \begin{tabularx}{0.975\textwidth}{l@{\extracolsep{\fill}}r r r r} 
  \toprule
Currency & Last timestamp & \#\,Blocks & \#\,Tx & \#\,Addresses \\ 
  \midrule
  BTC & 2025-03-31 23:33:50  & 890,325    & 1,173,115,412 & 1,386,674,802 \\ 
  LTC & 2025-03-31 23:58:44  & 2,871,702  &   316,376,996 &   340,449,985 \\ 
  DOGE & 2025-03-31 23:59:20 & 5,649,129  &   395,309,708 &   106,545,894 \\ 
  ZEC & 2025-03-31 23:59:10  & 2,873,734  &    15,285,542 &     8,424,388 \\ 
  ETH & 2025-03-31 23:59:59  & 22,170,335 & 2,749,975,760 &   340,621,351 \\ 
  TRX & 2025-03-31 23:59:57  & 70,937,424 & 9,690,710,588 &   324,382,918 \\ 
   \bottomrule
\end{tabularx}


  \caption{Analyzed cryptocurrencies with their respective last blocks used in
    the analysis.}
  \label{tab:blockchain-stats}
\end{table*}

\section{Cryptographic background}\label{sec:crypto_background}

\subsection{Public key formats}\label{subsec:key_formats}

In the analyzed ECDSA based cryptocurrencies, public keys are represented in
two primary formats, \emph{uncompressed} and \emph{compressed}. Each format
encodes a point $(x,y)$ on the elliptic curve \emph{secp256k1}.

\begin{description}
  \item[Uncompressed Public Key Format:] an uncompressed public key is a
    65-byte sequence. It begins with a \texttt{0x04} prefix, which explicitly
    denotes its uncompressed nature. The subsequent 32 bytes represent the
    $x$-coordinate of the elliptic curve point, followed by 32 bytes
    representing the $y$-coordinate.

  \item[Compressed Public Key Format:] a compressed public key is a more
    compact 33-byte representation, which starts with the prefix \texttt{0x02}
    or \texttt{0x03}. This initial byte serves as a parity indicator for the
    $y$-coordinate: \texttt{0x02} signifies an even $y$-coordinate, while
    \texttt{0x03} indicates an odd $y$-coordinate. The remaining 32 bytes
    exclusively encode the $x$-coordinate. To fully reconstruct the $(x,y)$
    point, the $y$-coordinate must be derived from the $x$-coordinate and its
    corresponding parity bit, utilizing modular arithmetic to solve the
    equation of the \emph{secp256k1} curve

\begin{equation}
  y^{2} \equiv x^{3} + 7 \pmod{p},
\end{equation}

where $p = 2^{256}-2^{32}-2^{9}-2^{8}-2^{7}-2^{6}-2^{4} - 1 $ is a large prime
number.
\end{description}

\subsection{Point on Curve Check}\label{subsec:point_on_curve}

To verify the extracted public keys in our dataset, we performed checks if the
retrieved public keys are indeed valid points on the elliptic curve
secp256k1\footnote{\url{https://gist.github.com/kernoelpanic/f6130da9bfbf387bc54209f5c6957a61}}.
For secp256k1, validating a public key means checking whether the point $Q$
lies on the curve and belongs to the correct subgroup. This involves first
decompressing the public key (if needed) and extracting the $(x, y)$
coordinates to verify that the point satisfies the curve equation
$y^2 \equiv x^3 + 7 \pmod{p}$, where $p$ is the prime field modulus
of secp256k1. Furthermore, it has to be ensured that the point is not the
point at infinity, $x$ and $y$ lie within the field $\mathbb{F}_p$, and
that $n \cdot Q = \mathcal{O}$, where $n$ is the order of the generator point
and $\mathcal{O}$ denotes the point at infinity. This final step ensures
subgroup membership \cite{secg2009sec1}.

When we are dealing with compressed public keys, there still is an
approximately $50\%$ chance that a random 32 byte sequence starting with
either \verb|0x02| or \verb|0x03| represents a valid point on the curve. This
is due to the fact that there are $(n - 1)/2 $ valid $x$ coordinates in
total, where $n$ is the total number of valid points on the curve. As each
$x$~coordinate has two valid $y$~coordinates ($y$ and $-y$) and one
point is the \emph{point-at-infinity}, we get $(n-1)/2$ valid $x$
coordinates. Therefore, the probability to hit a valid $x$ coordinate with a
random element in $\mathbb{F}_p$ is given by $(n - 1)/2 / p \approx 0.5$.

If extracting uncompressed keys (starting with \verb|0x04| followed by a 64
byte sequence representing the $x$ and $y$ coordinates) the probability
that a uniformly random pair $(x,y)$ results in a valid point on the curve can
be considered effectively zero for all practical purposes. As there are $n-1$
valid points on the curve in total (excluding the \emph{point-at-infinity})
and $p \cdot p$ possible pairs, the probability that a random pair
$x,y \in \mathbb{F}_p$ represents a valid point on the curve would be given by
$(n - 1) / p^2 \approx 2^{-256}$.

\section{UTXO transaction output types and public key extraction}\label{sec:utxo_types}

UTXO transaction outputs are categorized as follows in \emph{bitcoin-etl}:

\begin{description}
\item[Legacy Script Types]\leavevmode

  \begin{description}
    \item[\texttt{pubkeyhash} (Pay-to-Public-Key-Hash, P2PKH):] The
      historically original and most common script type. Funds are sent to a
      hash of a public key. Spending requires a digital signature from the
      corresponding private key. This is the format used by most standard
      Bitcoin addresses that start with the prefix~\texttt{1}.

    \item[\texttt{scripthash} (Pay-to-Script-Hash, P2SH):] This enables more
      complex transactions by sending funds to the hash of a script instead of a
      public key. The spender must provide the full script and a solution that
      satisfies it. It is commonly used for multi-signature wallets, with
      addresses typically starting with a 3. P2SH (Pay-to-Script-Hash) was
      enabled on the Bitcoin network at block 173,805, which was mined on
      April 1, 2012.

    \item[\texttt{nonstandard}:] This indicates a script that does not conform
      to any established patterns. Nodes generally do not relay or mine
      transactions with these outputs.
  \end{description}

\item[Segregated Witness Types] The SegWit upgrade separates witness data
  (signatures and scripts) from the transaction. As the witness data is stored
  separately, it is not part of the transaction ID (transaction hash) anymore
  resolving transaction malleability issues~\cite{bip141}. By discounting
  witness data, SegWit also increased the effective block size and thus
  transaction throughput. SegWit addresses typically begin with bc1.
  Segregated Witness (SegWit, BIP 141) was activated on the Bitcoin network
  on August 24, 2017, at block height 481,824. Litecoin was the first major
  cryptocurrency to activate SegWit (May 10, 2017).

  \begin{description}
    \item[\texttt{witness\_v0\_pubkeyhash} (Pay-to-Witness-Public-Key-Hash):]
    The SegWit equivalent of P2PKH is known as P2WPKH.

    \item[\texttt{witness\_v0\_scripthash} (Pay-to-Witness-Script-Hash):]
      Similarly, the native SegWit equivalent of P2SH for complex scripts is
      P2WSH.
    \item[\texttt{witness\_unknown}:] This is a forward-compatible type for
      future SegWit versions that are not yet recognized by the software.
  \end{description}

\item[Taproot]\leavevmode

  \begin{description}
    \item[\texttt{witness\_v1\_taproot} (Pay-to-Taproot, P2TR):] The most
      recent major protocol-level upgrade, which uses Schnorr signatures and
      Merkle trees to enhance privacy and efficiency~\cite{bip341}. This type
      makes complex spending conditions (like multi-signature or time-locked
      transactions) appear as simple spends on the blockchain. Taproot was
      activated on the Bitcoin network on November 14, 2021, at block
      height~709,632.
  \end{description}
\end{description}

\begin{figure*}[ht]
  \centering
  \includegraphics[width=0.975\textwidth]{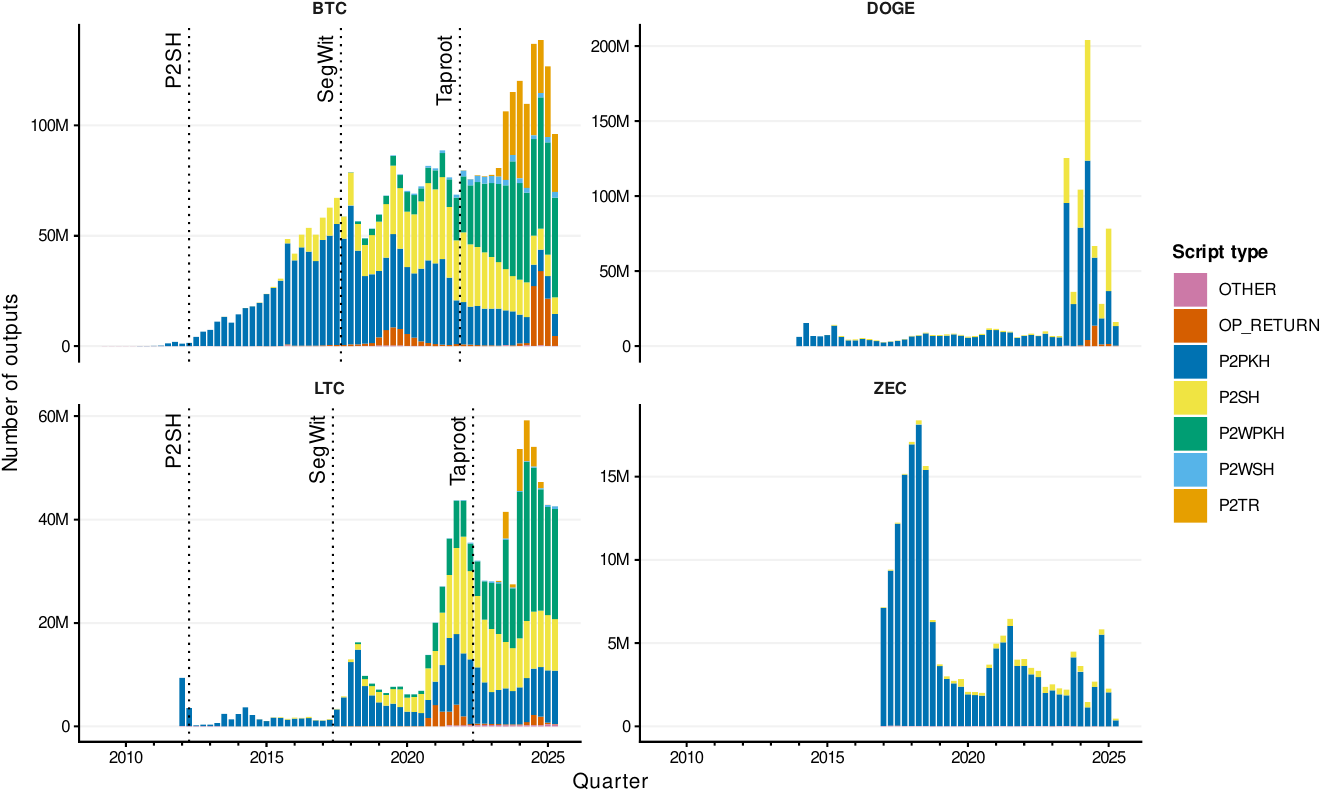}
  \caption{Quarterly counts of UTXO script types in transaction outputs.}
  \label{fig:utxo-script-types}
\end{figure*}

The quarterly counts of commonly used outputs are shown in Figure
\ref{fig:utxo-script-types} for the considered UTXO currencies. Different
blockchain services contribute to the BTC \texttt{OP\_RETURN} usage peak in
2019, most prevalently Veriblock (58\% of \texttt{OP\_RETURN} transactions) and
Omni/Tether (40\%) with a combined sum of 31.7M transactions between September
14, 2018 (the day of the first Veriblock transaction on Bitcoin), and December
31, 2019 \cite{strehle2020dominating}. The \texttt{OP\_RETURN} peak in Q2--Q4
of 2024 includes transactions of the Runes fungible token protocol, which was
launched on April 20, 2024 and generated an initial flurry of activity with
over 15~million transactions recorded in the first four months alone
\cite{cointel_runes2025}, with users' interest steadily declining towards the
end of 2024\footnote{https://www.theblock.co/data/on-chain-metrics/bitcoin/runes-transactions}.

\subsection{Public key extraction}\label{sec:utxo_extract}

The visibility and location of public keys within UTXO transaction scripts
vary depending on the script type, impacting privacy and data footprint on the
blockchain. In the following, we consider the characteristics of public key
exposure for the following script types:

\begin{description}

\item[P2PK:] For Pay-to-Public-Key outputs, the complete public key of the
  recipient is directly embedded within the locking script at the time of fund
  allocation. Consequently, the public key is immediately exposed on the
  blockchain upon transaction confirmation. Extraction of the public key can be
  performed by directly parsing the locking script, which is structured in the
  disassembled script assembly (\texttt{script\_asm)} representation as follows:

\begin{verbatim}
<Pubkey> OP_CHECKSIG
\end{verbatim}

\item[P2MS:] For \emph{bare} Pay-to-Multisig (BIP~11) outputs, the set of
  public keys involved in the multi-signature scheme is also directly situated
  within the locking script. This configuration specifies the total number of
  participating public keys ($n$) and the minimum number of required signatures
  ($m$) to authorize a spend. The \texttt{script\_asm} representation for a P2MS
  locking script adheres to the following format:

\begin{verbatim}
<m> <Pubkey 1> ... <Pubkey n> <n> OP_CHECKMULTISIG
\end{verbatim}

  P2MS is now largely considered non-standard and has been superseded by more
  advanced output types.

\item[P2PKH:] Pay-to-public-Key-Hash outputs embed only a cryptographic hash
  of the public key within their locking script. This design enhances privacy by
  not revealing the full public key until the associated funds are spent. When a
  UTXO is spent, the spender provides the public key and a digital signature
  within the unlocking (input) script of the transaction input.

\begin{verbatim}
<Signature> <Pubkey>
\end{verbatim}

  The provided public key gets hashed and compared against the hash of the
  public key stored in the locking script. If the hashes match and the signature
  is valid, the transaction is authorized.

\item[P2SH:] The Pay-to-Script-Hash transaction type was introduced to the
  Bitcoin protocol to enhance the flexibility and security of transaction
  outputs by allowing funds to be locked to the hash of a script rather than
  directly to a public key hash. This mechanism, specified in BIP 16, decouples
  the spending conditions from the initial transaction output, enabling the
  sender to commit to a complex spending script without needing its full
  details.

  The extraction of public keys from P2SH transactions necessitates an analysis
  of the \emph{redeem script} embedded within the signature script of the
  spending transaction.

  As all kinds of complex scripts are theoretically possible, we restrict the
  analysis to the following types:

\begin{description}
  \item[P2SH Multisignature Transactions:] P2SH is nowadays the prevailing
  method to implement multisig. The public keys are included in the redeem
  script, which adheres again to the structure

\begin{verbatim}
<m> <Pubkey 1> ... <Pubkey n> <n> OP_CHECKMULTISIG

\end{verbatim}

\item[P2SH-P2PKH] While technically feasible, the encapsulation of a
  traditional P2PKH script within a P2SH wrapper (P2SH-P2PKH) is not a standard
  or commonly implemented construction, as it offers no inherent advantages over
  a direct P2PKH transaction. The script signature has the structure

\begin{verbatim}
<Signature> <Pubkey> <Redeem script>

\end{verbatim}
\end{description}

\item[P2WPKH:] Pay-to-Witness-Public-Key-Hash represents an optimized version
  of P2PKH, introduced as part of \emph{Segregated Witness} (SegWit, BIP~141).
  Its locking script contains again only a cryptographic hash of the public key.
  The primary distinction from P2PKH lies in the placement of the unlocking
  information. The witness data provided in a separate part of the transaction
  serves as the unlocking mechanism, which has the same script structure as the
  P2PKH unlocking script.

\item[P2WSH:] Pay-to-Witness-Script-Hash operates on a principle similar to
  P2SH, enabling spending to the hash of a complex script. The key distinction
  lies in that the full witness script and its unlocking arguments are stored
  again in the separated witness data of the transaction.

\item[P2SH continued:] P2SH plays a crucial role in Bitcoin's evolution by
  enabling backward compatibility for SegWit transaction types, specifically
  P2WPKH and P2WSH. This nesting mechanism, often referred to as \emph{wrapped
  SegWit} allowed for the gradual adoption of SegWit benefits without requiring
  all wallets and services to immediately upgrade to support native SegWit
  addresses.

\begin{description}
\item[P2SH-P2WPKH:] When spending, the input's signature script contains the
  redeem script

\begin{verbatim}
OP_0 <20 byte pubkey hash>
\end{verbatim}

  The actual signature and public key are placed in the witness field, as they
  would be for a native P2WPKH spend

\begin{verbatim}
<Signature> <PubKey>
\end{verbatim}

\item[P2SH-P2WSH:] the redeem script is a specific SegWit version 0 P2WSH
  output script

\begin{verbatim}
OP_0 <32 byte witness script hash>
\end{verbatim}

The required data is again stored in the witness field analogous to P2WSH.
\end{description}
\end{description}

\section{Account based public key extraction}\label{sec:account_based_extract}

Ethereum and Tron, which is conceptually based on Ethereum\footnote{Tron also
supports smart contracts through their TVM (Tron Virtual Machine) which is
based on the EVM (Ethereum Virtual Machine).}, use the account model initially
introduced by Ethereum. In this account model, the public key used to generate
an account address is not transmitted explicitly with transactions. Instead, a
recovery process is used that enables the reconstruction of the signer's
public key from the \emph{signature} and the \emph{message hash}. Thereby,
transaction authenticity can be verified without requiring additional public
key data within a transaction~\cite{wood2014ethereum}.

The method is based on the properties of the \emph{secp256k1} elliptic curve,
which allow recovering the ECDSA public key from an ECDSA signature using the
public key recovery operation described in~\cite{secg2009sec1}. ECDSA
signatures consist of the tuple $(r, s, v)$, where $r$ and $s$ are standard
ECDSA signature components and $v$ serves as an optional \emph{recovery
identifier} for faster public key recovery. Alternatively, all recoverable
points can be tried, which are $2 (h+1)$, where $h$ is the cofactor of the
underlying elliptic curve. Thus, for \emph{secp256k1} where $h = 1$ the value
$v$ can be 0, 1, 2 or 3, where the first two values are most
common\footnote{As $r$ represents the $x$ coordinate in $\mathbb{F}_p$, $x$
can be $r$ or $r + n$, and both $x$ coordinates can have $y$ or $-y$ (i.e.,
$p - y$) leading to a valid point on the curve, although $x$ values where the
magnitude is greater than the curve order are uncommon.}.

In context of cryptocurrencies the byte representing $v$ might encode
additional information. For example, indicating a compressed or uncompressed
key format in case of Bitcoin, or in case of Ethereum $v$ encodes the chain ID
as a replay protection mechanism due to the Ethereum Classic
hard fork~\cite{Buterin2016EIP155}. Therefore, in Ethereum $v \in \{
\text{chainId} \cdot 2 + 35, \text{chainId} \cdot 2 + 36\}$. Due to these
overloaded custom encodings, common libraries used for public key recovery in
\emph{secp256k1} might not know how to decode $v$ correctly and thus throw an
error when given a message signature pair with a custom encoding as $v$ value.
Therefore, either $v$ is manually replaced by the appropriate value, or all
four recoverable points are tried and checked against the source address.

Moreover, in the case of Ethereum, special care has to be taken how the
Keccak-256 hash representing the message is computed in different Ethereum
transaction types, as the hash that is signed using ECDSA is also not directly
part of the transaction data structure returned by common Ethereum
APIs\footnote{\url{https://gist.github.com/kernoelpanic/423c61f90e81e4c9d473ff6fda783559}}.

\begin{description}
\item[Ethereum:] For our study we went over all Ethereum transactions
  performed by EOAs and extracted \numUniqueEth\ unique public keys from
  Ethereum transactions using the previously described public key recovery
  method, as the public key is not available via some API directly.
\item[Tron:] Due to the large number of Tron transactions we did not perform
  signature recovery for every single transaction and instead first identified
  the subset of 266,333,226 transactions where a new source/owner address has
  performed its first transaction before recovering the unique associated public
  keys from the respective transactions.
\end{description}

For the account based public key extraction in both Ethereum and Tron we
limited ourselves to the public keys directly associated with EOAs. Because of
the smart contract functionality in both systems it is also possible to
perform signature verification within the contract execution in the virtual
machine, e.g. on signed messages that are provided as input data. Further,
Tron introduces the concept of permissions to accounts that allow additional
signers to be specified next to the owner\footnote{cf.
\url{https://developers.tron.network/docs/multi-signature}}. We leave the
extraction and analysis of this keying material to future work.

\section{Detailed method validation procedure}\label{sec:validation_procedure}

Since extracting public keys requires active usage of it to sign at least one
transaction as a sender, our method to identify key reuse is limited to such
instances where a key has been used at least once. Therefore, we are not able
to identify key reuse across different address formats and possibly even
cryptocurrencies if the underlying public key has never been actively used to
sign a transaction. This scenario, which we call \emph{exclusively passive}
(re)use, cannot be detected by our approach. In the following we try
to estimate the amount of \emph{exclusively passive} reuse by evaluating
special cases in which such events can be detected by other means and
comparing it to the results we get through our public key based approach.

\subsection{Bitcoin, Litecoin, Dogecoin address comparison}\label{subsec:bitcoin_litecoin_doge_validation}

The approach of HSI for comparing addresses between different Bitcoin-like
UTXO cryptocurrencies relies on the fact that different address formats (see
Appendix \ref{sec:utxo_extract}) encode the hash of the underlying public key
or unlock script that is required to spend the output. Therefore, the
underlying hash in the UTXO is compared and not the encoded address itself.
The advantage of this approach: It can also identify reuse of addresses (and
thus keys) which are used \emph{exclusively passive}. The disadvantage of the
hash based approach is that this technique is only applicable for the same
underlying hash formats, such as Bitcoin's RIPEMD160 representation, and not
across UTXO and account based systems. For example, P2PKH uses the same
underlying RIPEMD160 representation of the public key across Bitcoin, Litecoin
and Dogecoin. By extracting this hash from the output script, the overlap
between those cryptocurrencies can be computed, even if the respective UTXO
output has never been spent. The same can also hold true for different address
types, such as P2PKH and P2WPKH, but not for Ethereum based addresses. By
running our public key based analysis for the block range in HSI, we are able
to directly compare both approaches against each other. For an overview see
Table~\ref{tab:harrigan}.
As expected, by relying on public keys, instances of \emph{exclusively
passive} cross-chain reuse cannot be identified. Therefore, we identified
\numPercentHarrignBtcLtc\% of the cases for BTC-LTC, while for DOGE-BTC we
identified \numPercentHarrignBtcDoge\% of the cases and
\numPercentHarrignDogeLtc\% for DOGE-LTC. This number can be increased by
taking all discovered public keys, including the ones first revealed after the
block range considered in HSI, into account. This indicates that the public
key based method gets better over time as more and more public keys get
revealed. Surprisingly, when recomputing the numbers according to HSI's
methodology for our current block range, our public key based approach exceeds
expectations, ranging from \numPercentHarrignAllLtcBtc\% for BTC-LTC to
\numPercentHarrignAllBtcDoge\% for BTC-DOGE and \numPercentHarrignAllLtcDoge\%
for LTC-DOGE. Possible explanations for these high numbers are the
incomparability of P2WSH, which uses SHA256, compared to P2SH which relies on
RIPEMD160, or the utilization of both compressed and uncompressed keys which
would also result in different hashes and can only be detected through our key
based approach. Moreover, Dogecoin makes extensive use of P2SH, for which
address based reuse checks are less effective, e.g., in case of multi-sig
transaction, compared to public key extraction.

\begin{table}[ht]
\centering
\begin{tabular}{l|r|r|r|c|>{\columncolor{gray!20}}r|>{\columncolor{gray!20}}c}
\thead{\textbf{Reuse} \\ \textbf{between}} &
\thead{\textbf{Block} \textbf{height}} &
\thead{\textbf{Reuse} \\ \textbf{HSI}} &
\thead{\textbf{Reuse} \\ \textbf{Public key}} &
\thead{\textbf{Reuse} \\ \textbf{\% HSI}} &
\thead{\textbf{Reuse} \\ \textbf{Public key (all)}} &
\thead{\textbf{Reuse} \\ \textbf{\% HSI}} \\
\hline\hline
\thead{Bitcoin \\ Litecoin} &
\thead{\numBtcHeightHarrigan \\ \numLtcHeightHarrigan} &
\numReuseHarrignBtcLtc &
\numReusePkBtcLtc &
\numPercentHarrignBtcLtc\% &
\numReusePkBtcLtcAll &
\numPercentHarrignBtcLtcAll\%
\\
\hline
\thead{Bitcoin \\ Litecoin} &
\thead{\numBtcHeight \\ \numLtcHeight} &
\numReuseHarrignAllLtcBtc &
\numReusePkAllLtcBtc &
\numPercentHarrignAllLtcBtc\% &
- &
-
\\
\hline\hline
\thead{Bitcoin \\ Dogecoin} &
\thead{\numBtcHeightHarrigan \\ \numDogeHeightHarrigan} &
\numReuseHarrignBtcDoge &
\numReusePkBtcDoge &
\numPercentHarrignBtcDoge\% &
\numReusePkBtcDogeAll &
\numPercentHarrignBtcDogeAll\%
\\
\hline
\thead{Bitcoin \\ Dogecoin} &
\thead{\numBtcHeight \\ \numDogeHeight} &
\numReuseHarrignAllBtcDoge &
\numReusePkAllBtcDoge &
\numPercentHarrignAllBtcDoge\% &
- &
-
\\
\hline\hline
\thead{Litecoin \\ Dogecoin} &
\thead{\numLtcHeightHarrigan \\ \numDogeHeightHarrigan} &
\numReuseHarrignDogeLtc &
\numReusePkDogeLtc &
\numPercentHarrignDogeLtc\% &
\numReusePkDogeLtcAll &
\numPercentHarrignDogeLtcAll\%
\\
\hline
\thead{Litecoin \\ Dogecoin} &
\thead{\numLtcHeight \\ \numDogeHeight} &
\numReuseHarrignAllLtcDoge &
\numReusePkAllLtcDoge &
\numPercentHarrignAllLtcDoge\% &
- &
-
\\
\hline\hline
\thead{Bitcoin \\ Litecoin \\ Dogecoin} & \thead{\numBtcHeightHarrigan \\ \numLtcHeightHarrigan \\ \numDogeHeightHarrigan} &
\numReuseHarrignBtcLtcDoge &
\numReusePkBtcLtcDoge &
\numPercentHarrignBtcLtcDoge\% &
\numReusePkBtcLtcDogeAll &
\numPercentHarrignBtcLtcDogeAll\%
\\
\hline
\thead{Bitcoin \\ Litecoin \\ Dogecoin} & \thead{\numBtcHeight \\ \numLtcHeight \\ \numDogeHeight} &
\numReuseHarrignAllLtcBtcDoge &
\numReusePkAllLtcBtcDoge &
\numPercentHarrignAllLtcBtcDoge\% &
- &
-
\\
\hline
\end{tabular}
\caption{Comparison of analyzed cryptocurrencies with their respective reused
  address hashes (according to HSI~\cite{harrigan2018airdrops}) and keys
  (according to the paper at hand) as well as the last blocks used in the
  analysis.
  The last two gray columns represent our public key based method applied to
  the block range of HSI, but taking all public keys into consideration that
  have since been revealed in the respective chains. In the case of
  BTC/LTC/DOGE we are actually able to surpass the address reuse based result.
  This is because the key based analysis includes and considers reuse between
  both compressed and uncompressed public keys, which can not be detected
  through address reuse alone.}
\label{tab:harrigan}
\end{table}

\subsection{Ethereum-Tron address comparison}\label{sec:ethereum_tron_validation}

The Tron address format essentially represents a different \emph{encoding} of
the address format used in Ethereum and most EVM-compatible cryptocurrencies.
Therefore, both address types can be converted into each other without
knowledge of the underlying public key. To obtain the corresponding Tron
address from an EVM-compatible address a \verb|0x41| prefix and subsequently a
4 Byte checksum is added and the resulting hex string converted to Base58
encoding (i.e., Base58Check)~\cite{tron:accounts-doc}.

Hence, Tron addresses can also be converted back to Ethereum addresses by
decoding the Base58Check format back to hexadecimal and removing the
\verb|0x41| prefix. This prefix is always set so that all Tron addresses start
with a \verb|T|, e.g., \texttt{\textcolor{red}{T}ErNaNKA\dots{}MKJb26A9wn}
which decodes to \texttt{0x\textcolor{red}{41}358DB59BC1\dots{}71F19FAC8} and
thus is also a valid Ethereum address after removing the prefix byte and
checksum.

This interchangeable address format can be used to validate our approach of
identifying key reuse and associated addresses by starting with (recovered)
elliptic curve public keys. Given the address format similarities of Tron and
Ethereum we are able to identify key reuse across those cryptocurrencies even
if the respective key has never actually been used to sign a transaction in
any of the two cryptocurrencies.

By converting all Tron and Ethereum addresses to a common 40 character (20
byte) hex format, we were able to identify the overall key reuse across both
cryptocurrencies, which amounts to \numTrxEthPassive~unique
keys\footnote{Representing between \numTrxEthPassivePercentofAddresses\% and
\numTrxEthPassivePercentofPublicKeys\% of all unique addresses/public keys in
Ethereum.}. Compared to that, \numTrxEthActive~of those keys have been
actively reused in one of the two cryptocurrencies. Thus those
\emph{actively} used keys would also have been found by our approach that
extracts the respective public keys of the sender. The difference of
\numTrxEthDelta~addresses is the \emph{exclusively passive} key reuse (i.e.,
only passive key usage in both chains) we would not have been able to identify
with our approach. In other words, our approach which relies on active usage
of an underlying key-pair for sending/signing a transaction at least once would
have detected \numTrxEthActivePercent\% of reused keys in the case of Tron
and Ethereum. Of course it is difficult to generalize this to other
cryptocurrency combinations, but the high detection rate is a good indicator
of the general feasibility of the approach.

To rule out false positives arising from overlapping smart contract addresses,
we looked at the source code regarding how smart contract addresses are
generated in Tron~\cite{tron-java-walletutil-line} and
Ethereum~\cite{go-ethereum-crypto-line,go-ethereum-crypto-doc} and found that
the address formats are different and thus cannot lead to overlapping contract
addresses. Note that this might not be true for other combinations of EVM
compatible cryptocurrencies~\cite{axelar2024sameaddress} (we refer to the
discussion in Section~\ref{sec:discussion} regarding this practice).

\section{Active key reuse}\label{sec:appendix_active_reuse}

The full distribution of active public key reuse across the different sets is
visualized in the UpSet plot in Figure~\ref{fig:upset-plot-full}.

\begin{landscape}
  \topskip0pt
  \vspace*{\fill}
  \begin{figure*}[ht]
    \centering
    \includegraphics[scale=0.3]{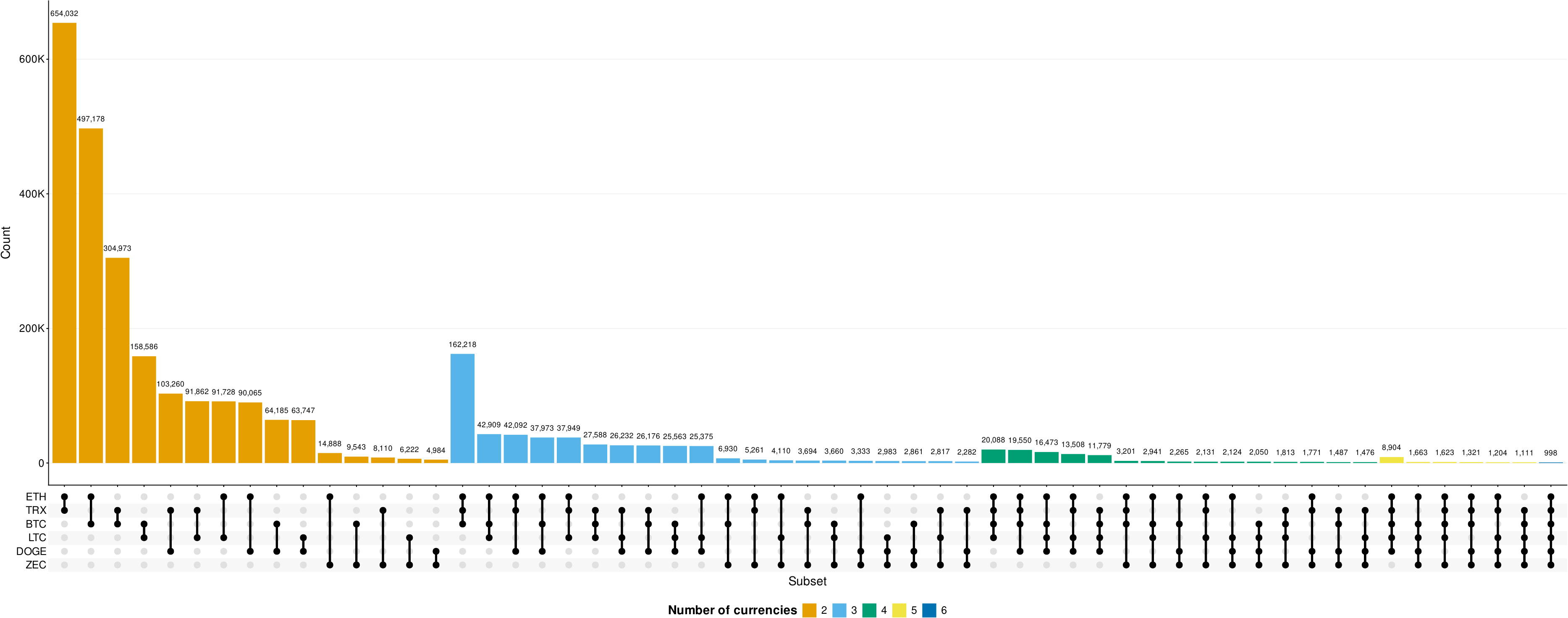}
    \caption{Cardinality of all set intersections for active public key reuse
      across the considered currencies.}
    \label{fig:upset-plot-full}
  \end{figure*}
  \vspace*{\fill}
\end{landscape}

\section{Improve available clustering using reused keys}\label{sec:improve_clustering}

\begin{figure*}[ht]
  \centering
  \includegraphics[width=0.5\textwidth]{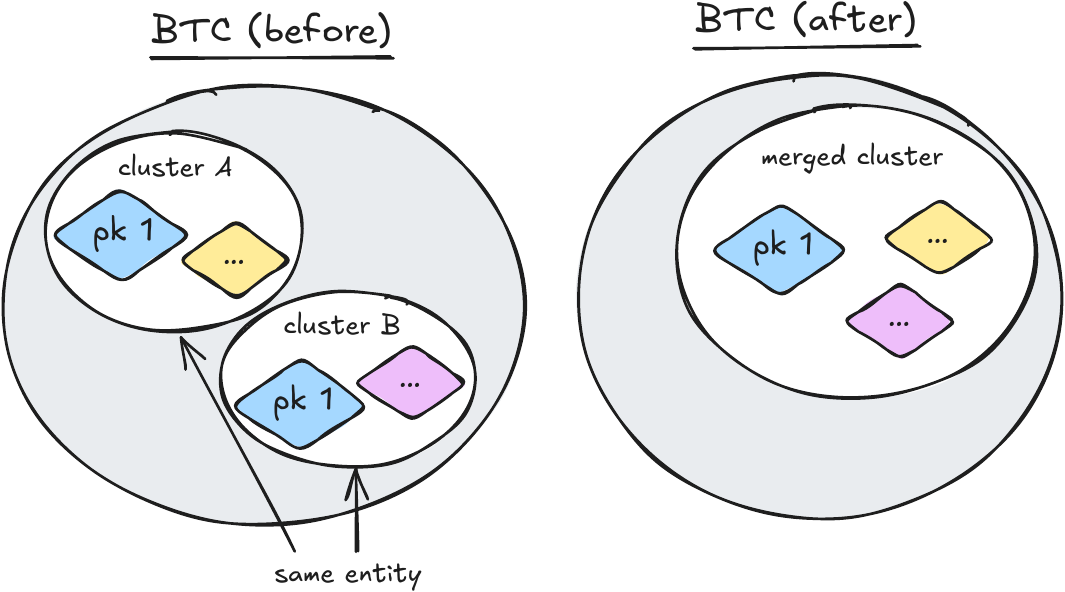}
  \caption{Depiction of how any available clustering on address level, such as
    multiple-input clustering in UTXO based cryptocurrencies, can be improved by
    deriving different addresses from the same underlying public key
    (\emph{pk 1} in the figure). If the derived addresses are in different
    clusters, the clusters can be merged.}
  \label{fig:cluster_int}
\end{figure*}

\section{Method for Identifying HD Wallets through cross-chain Key Reuse}\label{app:hd-method}

One possible (weak) indicator of a reused HD scheme across two different UTXO
cryptocurrencies $A$ and $B$ could be obtained as follows:

If we observe a transaction originating from a reused key, which we identify
as an input, and a reused key as output in cryptocurrency $A$, as well as the
\emph{same} reused key as input and the \emph{same} reused key as output in
cryptocurrency $B$, then those reused keys could have been derived via the
same HD scheme in both currencies, where the outputs are newly generated
change addresses.

\section{Clustering of Tornado Cash Recipients through cross-chain Key Reuse}\label{sec:tornadocash}

In the following we briefly highlight the possible privacy implications of
unintentional cross-chain key reuse in the context of mixing services and
anonymization techniques. Hereby, we identify a DeFi Bridge (Avalanche Bridge)
as the potential culprit for a large number of key reuse events that we also
associate with Tornado Cash interactions.

In Subsection~\ref{subsec:entities_reuse} we outline that we were able to
match 203 keys from our dataset of reused keys to mixing services using
labels from Dune Analytics, specifically in this case all of the 203 matches
relate to associations with \emph{Tornado Cash}. As outlined in the Tornado
Cash whitepaper\footnote{cf.\ %
\url{https://web.archive.org/web/20211026053443/https://tornado.cash/audits/TornadoCash_whitepaper_v1.4.pdf}},
it \enquote{\ldots{} implements an Ethereum zero-knowledge privacy solution: a smart
contract that accepts transactions in Ether (in future also in ERC-20 tokens)
so that the amount can be later withdrawn with no reference to the original
transaction.}

Hereby, a crucial component to ensure privacy is the ability to withdraw funds
in a privacy preserving manner without allowing for associations with the
initial deposit transaction(s)~\cite{tang2021analysis}. If the user of such a
service inadvertently has performed (cross-chain) key reuse with the receiving
key, it may again become possible to associate the withdrawal address with the
original depositing entity.

To evaluate the feasibility of this approach, we perform a manual explorative
analysis of a subset of the 203 public keys we associate with the Tornado Cash
mixing service where we could identify key reuse, specifically between Bitcoin
and Ethereum. Hereby, we first investigate the transaction history of the
associated Ethereum accounts and subsequently explore the transaction patterns
for the reused keys in Bitcoin.

While many of the sampled accounts exhibit patterns in Ethereum that already
appear unsuitable to preserve privacy, e.g., where non-blinded incoming
deposit transactions were also performed, we are also able to identify several
accounts that are initially funded through Tornado Cash and appear to uphold
strict separation. However, by observing the corresponding cross-chain key
reuse transactions in Bitcoin, additional information on the entity may be
learned in either case, leading to a further loss of privacy. We therefore
also refrain from providing concrete examples from our dataset as it is
unclear if this could negatively impact the affected individuals.

An interesting and worrying observation we make in this context is that a
large number of the Ethereum accounts in our Tornado Cash tagged set, which
exhibit key reuse in Bitcoin, have also reused their public key (address) in
the EVM compatible Avalanche network. An in-depth manual analysis of select
examples reveals that the corresponding reused addresses in Bitcoin were
funded from address \texttt{bc1q2f0tczgrukdxjrhhadpft2fehzpcrwrz549u90}, which
corresponds to the Avalanche Bridge Bitcoin address\footnote{cf.\ %
\url{https://support.avax.network/en/articles/6349706-what-are-the-avalanche-bridge-addresses-on-the-bitcoin-and-avalanche-networks}}.
Upon further investigation, we find that the design of the Intel SGX-based
Avalanche Bridge service is built such that Bitcoin addresses are derived from
the public key recovered from the Avalanche transaction that burns the wrapped
asset\footnote{cf.\ %
\url{https://medium.com/avalancheavax/bridging-bitcoin-to-avalanche-a-technical-overview-2535e7088b8}}.

In the case of the affected parties in our Tornado Cash dataset, additional
key reuse between Avalanche and Ethereum has also taken place, which appears
to be a prevalent phenomenon in EVM-compatible designs (see
Section~\ref{sec:background}). We therefore were able to associate these
Tornado Cash funded accounts in Ethereum with UTXO transactions in Bitcoin,
even though our original key reuse analysis did not include the Avalanche
network, where the swap to Bitcoin took place.

Using this gained insight, we revisit the Bitcoin component of our full
dataset of reused keys in order to identify funding transactions that were
initiated from the Avalanche Bridge, and are able to identify
\numReusedTxnsAVAX{} transactions involving \numReusedKeysAVAX{} unique keys.
This example demonstrates that the usability design decision to reuse keys in
the Avalanche Bridge directly negatively impacts user privacy by enabling the
linkage between depositing and withdrawing transactions of these cross-chain
transfers across multiple cryptocurrency networks. It is unclear if the users
of the Avalanche Bridge are aware of the implicit key reuse that takes place
when engaging with this service.

\end{document}